\documentclass[preprint,aps]{revtex4}
\usepackage{graphicx}
\usepackage{dcolumn}
\usepackage{bm}

\begin{document}
\title{On/off switching of adhesion in gecko-inspired adhesives}

\author{Tetsuo Yamaguchi$^{1}$, 
        Akira Akamine$^{2}$ 
        and Yoshinori Sawae$^{2,3}$}

\affiliation{
1 Department of Biomaterial Sciences, 
The University of Tokyo, Tokyo 113-8657 Japan\\
2 Department of Mechanical Engineering, 
Kyushu University, Fukuoka 819-0395 Japan\\ 
3 International Institute for Carbon-Neutral Energy Research, 
Kyushu University, Fukuoka 819-0395 Japan
}


\date{\today}

\begin{abstract}
In this study, the adhesion-detachment behaviour of a gecko-inspired adhesive pad 
was investigated to understand the on/off switching mechanisms of adhesion in gecko feet. 
A macroscopic spatula model was fabricated using silicone rubber, and adhesion tests 
combining lateral sliding and vertical debonding were conducted. 
It was observed that the contact state and the adhesion force of the pad vary 
considerably with the direction of lateral sliding prior to debonding, 
and that the pad achieves adhesion during debonding even when it loses contact 
due to excess lateral sliding. 
These results explain the mechanisms behind the on/off switching 
and stable adhesion of gecko feet, and suggest the possibility of 
developing new-generation adhesives capable of switchable adhesion. 
\end{abstract}

\maketitle

\section{Introduction}
Geckos are able not only to walk on the ground, but also 
to climb up walls and to hang upside-down on ceilings\cite{Maderson}-\cite{Autumn1}. 
As geckos walk under a small or negative normal load in such situations, 
they must attach their feet onto a substrate firmly, 
and detach from the substrate quickly\cite{Autumn1}-\cite{Hagey}.\\ 

Microscopic observations and adhesion experiments involving the fine structures 
of gecko feet have revealed that van der Waals attractive forces 
between the substrate and the pad at the edge of the feet 
play a major role in adhesion\cite{Autumn4}, 
similar to other conventional adhesive systems\cite{Pocius}. 
Moreover, different aspects, 
such as a slender geometry and asymmetric structures, 
contribute to quick adhesion and detachment\cite{Ruibal}. 
In fact, many studies have regarded gecko feet as microscopic arrays 
of fine hairs, and researchers have fabricated gecko-inspired adhesives 
with fibres or pillars on the surface\cite{Geim}-\cite{Wang}.\\

In addition, recent theoretical studies\cite{Zhao}-\cite{Gillies} 
have highlighted the importance of the geometry with a tilted pad attached 
to a curved beam and have successfully explained some of the superior properties 
of gecko feet. In particular, Zhao and co-workers\cite{Zhao} discussed that 
in such a structure, the adhesion of the pad depends strongly on the operation 
conditions before debonding, i.e., strong adhesion is achieved 
when the pad is pulled laterally before vertical debonding, 
and quick release is realized when the pad is pushed laterally 
before the debonding. With these lateral operations in addition 
to the vertical debonding, it was expected that on/off switching 
of the adhesion can be easily performed. 
Furthermore, Yamaguchi and co-workers\cite{Yamaguchi} showed 
that the adhesive pad geometry of gecko feet contributes 
to the generation of a negative normal force during sliding friction 
and yields stable motion without unexpected detachment from the substrate; 
this is caused by repetitive stick-slip motions and 
the coupling between the vertical and horizontal forces 
of a curved beam. Bio-inspired designs resembling gecko feet in terms of 
adhesion and friction are considered to be important for creating 
bio-inspired, stimuli-responsive, soft interfacial devices\cite{Spolenak}-\cite{Croll}.\\

However, micro-mechanical studies involving adhesion experiments and 
in-situ visualization for the development of gecko-feet-like asymmetric 
slender structures remain inadequate. Although several trials have been conducted thus far, 
it remains difficult to fabricate arrays of fine and well-controlled structures 
having the aforementioned geometry\cite{Murphy}.\\

In this study, we fabricated a macroscopic analogue of a gecko-inspired tilted adhesive pad 
attached to a curved beam, and observed the adhesion/detachment behaviour 
of a single pad. During adhesion testing, we combined vertical compression-debonding 
motions with lateral sliding motions. This type of testing geometry is known as 
frictional adhesion\cite{Autumn1, Autumn3, Hagey, Yamaguchi},\cite{Gravish}-\cite{Tian}, 
in which the coupling between adhesion and friction is important, and materials 
with strong adhesion and high friction are normally used\cite{Yamaguchi3}. 
As discussed subsequently, the adhesion properties strongly depend on the precursory 
lateral sliding motions, and on/off switching of adhesion is possible 
with lateral sliding in addition to vertical debonding.\\

\section{Experiment}
\subsection{Sample}
A schematic of a sample is depicted in Fig. 1(a). 
The sample, having the structure of a single curved beam 
with a tilted pad on the tip, was prepared 
by curing silicone prepolymer (SILPOT184 prepolymer, TORAY DOW CORNING, Japan) 
with a curing agent (SILPOT184 CAT) in a Plexiglass mould, 
with a weight ratio of 9:1 at $90^{\circ}{\kern-0.5mm}$C for 5 h.\\

\begin{figure}
\includegraphics[width = 80mm]{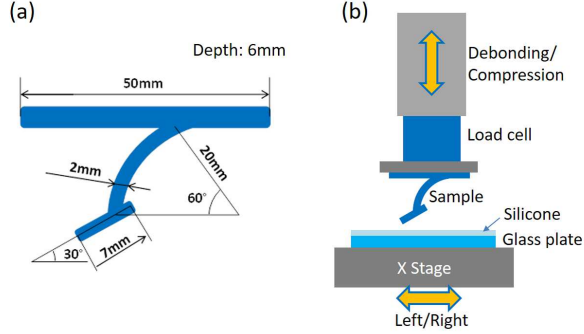}
\caption{Schematic of (a) sample and (b) experimental setup.}
\label{fig1}
\end{figure}

\subsection{Substrate}
In this study, we aimed to reproduce realistic situations involving microscopic gecko 
pads and to understand the underlying mechanisms using a macroscopic analogue 
(our adhesive pad model). Here, we considered the following assumptions: 
\begin{itemize}
\item The adhesion energy $E_{adhesion}$ and the stored elastic energy $E_{elastic}$ 
are the main factors controlling the functioning of the pad.\\
\item The deformation of a microscopic gecko pad (and a spatula) and a macroscopic analogue model 
are similar to each other (i.e., the strains are the same in the two systems). 
\end{itemize}

Based on these assumptions, we can deduce that the adhesion energy is proportional to $L^2$ 
(where $L$ is the system size), while the stored elastic energy is proportional to $L^3$. 
Furthermore, to reproduce the same mechanics between gecko feet and our model, we assumed 
that the ratio between the adhesion energy and the stored elastic energy is the same 
in gecko feet and our model. 

\begin{eqnarray}
	\frac{E_{adhesion}}{E_{elstic}} &\propto& \frac{\gamma}{E L} \nonumber \\
	                                &\sim& \frac{ 10^{-2} J }{ 10^9 Pa \times 10^{-6} m } \hspace{5mm} (gecko) \nonumber \\
	                                &\sim& \frac{ \gamma_{model} }{ 10^{6} Pa \times 10^{-1} m } \hspace{5mm} (model), 
	\label{eq1}
\end{eqnarray}

where $\gamma$ is the surface energy between a pad and a substrate, and $E$ is the Young's modulus 
of the pad. The values for gecko feet were obtained from the literature\cite{Gravish} and those for our model were estimated 
through rheological measurements. Based on Eq. (\ref{eq1}), the surface energy must be as large as 
$\gamma_{model} \sim 1 J$, which cannot be reached as long as the original glass surface is used ($\sim 10^{-2}-10^{-1} J$). 
To enhance the stickiness of the glass, we coated silicone rubber with 3wt\% of a curing agent (SILPOT184, TORAY DOW CORNING, Japan) 
and cured it at $90^{\circ}{\kern-0.5mm}$C for 5 h.\\

\subsection{Adhesion test}
The experimental setup for the adhesion test is illustrated in Fig. 1(b). 
The sample was attached to the head of the tensile testing machine (MST-1, Shimadzu, Japan), 
and the substrate was fixed onto the horizontally moving stage (SGSP80-20ZF, Sigma-Koki, Japan).\\

In the first series of experiments, we examined the on/off switching of the adhesion properties. 
To achieve this, we combined lateral motions with compression/tension during the adhesion test, 
as illustrated in Fig. 2. Before the experiment, we separated the sample from the substrate by 2.5 mm. 
Then, we applied a compressive force on the sample in the vertical direction 
for 5 s at 2 mm/s; thus, the compression displacement after contact was 7.5 mm. 
After the compression, we translated the substrate either to the left 
or to the right at 1 mm/s or maintained the position (at 0 mm/s) for 10 s. 
Thereafter, we pulled the sample in the vertical direction at 2 mm/s until 
the sample detached from the substrate completely. 
The vertical tensile force was measured as an adhesion force using a load cell.\\

\begin{figure}
\includegraphics[width = 80mm]{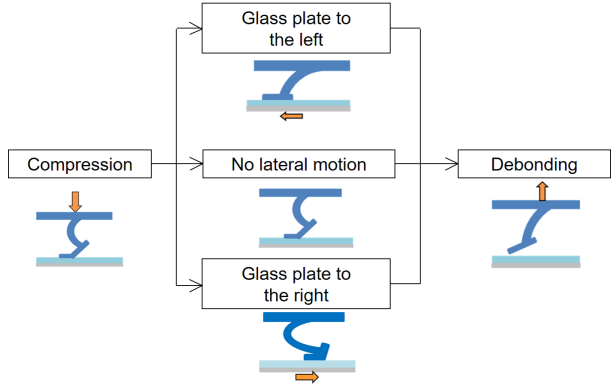}
\caption{Three different protocols for the first series of adhesion tests.}
\label{fig2}
\end{figure}

In the second series of experiments, we investigated how excess 
sliding and vertical compression affect the debonding behaviour: 
after applying compression at 2 mm/s, we translated the bottom substrate 
only to the left by 20 mm; then, we pulled the sample 
in the vertical direction at 2 mm/s until complete detachment, 
in the same manner as in the first series of the experiments. 
However, in the second series of experiments, 
we controlled the vertical compression displacement from 4 to 8 mm and 
changed the lateral sliding speed from 0.01 to 1 mm/s.\\

\section{Results and discussion}
\subsection{On/off switching behaviour}
Fig. 3 shows the time evolutions of the vertical tensile forces for 
the three different testing protocols in the first series of the experiments. 
When the substrate was translated to the right (R) or was kept at the same position (Keep) 
in the lateral sliding phase, no tensile force was generated during the final debonding phase. 
In contrast, when the substrate was translated to the left (L), a tensile force was generated 
during the final debonding phase. These results clearly show that the generation 
of the tensile force depends strongly on the precursory lateral sliding motions.\\

To explain the mechanisms behind the large differences in the debonding force, 
we show snapshots of the deformation states in the right panel of Fig. 3. 
When the compression was applied (A), the entire surface of the pad did not attach 
to the substrate; only the toe did. The reason for this imperfect contact is 
that the pad rotates in the counter-clockwise direction 
due to the buckling of the stem (curved beam). 
However, the differences in the deformation state appeared after the lateral sliding: 
as the substrate was translated to the left, the pad started to rotate in the 
clockwise direction, and the entire surface of the pad became attached, as shown in B. 
Once perfect contact was attained, a strong vertical tensile force was generated, 
and this force detached the pad from the substrate, as shown in the left panel of Fig. 3. 
In contrast, when the substrate was translated to the right or when 
it was kept at the same position during the lateral sliding phase, the contact states 
immediately before debonding were C and A respectively, and almost no tensile force was generated. 
In both cases, the imperfect contact produced an edge in the contact 
region, and this edge caused a stress concentration during debonding. This resulted in 
the weak adhesion force.\\

These results suggest that geckos achieve on/off switching of the adhesive force 
by performing translation in the horizontal direction, in addition to the vertical direction.\\

\begin{figure}
\includegraphics[width = 80mm]{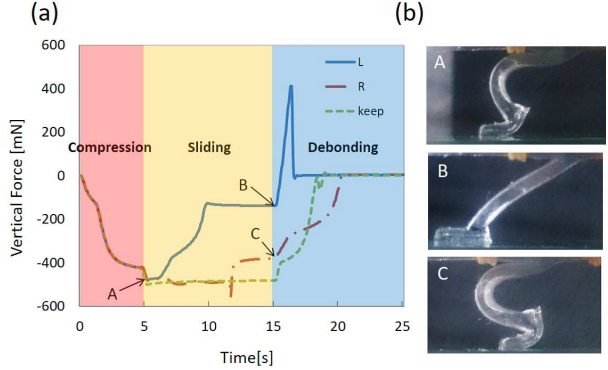}
\caption{Left: Time evolutions of the vertical forces for three different testing 
protocols. Right: Snapshots depicting the deformation of the samples. 
Letters A--C in the left panel correspond to those in the right panel.}
\label{fig3}
\end{figure}

\subsection{Effects of compression displacement and sliding distance}
As discussed above, the lateral sliding produces additional rotation of the pad 
and controls the contact state. However,
the aforementioned setup was special in terms of the vertical compression displacement 
and lateral sliding distance, which were 7.5 mm and 10 mm, respectively. 
In actual situations involving gecko feet, this corresponds to 
fine control of the displacements on a micron scale, 
which founds to be impossible or difficult for geckos.\\

We investigated the effects of excess vertical compression and 
lateral sliding. In the second set of the experiments, we changed 
the compression displacement from 4 to 8 mm and set the sliding 
distance as 20 mm, which is the twice that in the on/off switching experiment. 
The results are illustrated in Fig. 4. Here, the differences are coming 
from the differences in the compression displacement.\\

When the applied compression displacement was 4 mm (Fig. 4(a)), 
almost no tensile force was generated during lateral sliding, except 
at the early stage. The mechanism behind the generation of the weak force 
was that the pad formed partial contact only at the toe, 
as shown in the right panel of Fig. 4(a). 
In contrast, when the compression displacement was 6 mm (see Fig. 4(b)), 
the situation changed drastically: a strong adhesion force was generated 
at the initial phase of the lateral sliding, and adhesion forces were successively 
generated during the subsequent phases. 
This was due to the near-perfect contact between the pad and substrate during the lateral 
sliding, as seen in the right panel of Fig. 4(b). 
However, when the compression displacement was 7.5 mm (Fig. 4(c)), 
A tensile force was generated at the early stage, and almost no force 
was generated thereafter; this is similar to the case of the small compression displacement 
(4 mm). However, an adhesion force was again generated in the final debonding phase. 
This can be explained by the snapshots in Fig. 4; during the lateral sliding, 
the heel of the pad attained contact, but in the debonding phase, 
the pad rotated in the counter-clockwise direction and formed perfect contact 
before complete debonding was achieved.\\

These behaviours are similar to the sliding friction of a block against a compliant gel\cite{Yamaguchi3}; 
The angle of the inclined block strongly affects the contact state and stick-slip 
motions, due to the bi-material effect.\\

\begin{figure}
\includegraphics[width = 80mm]{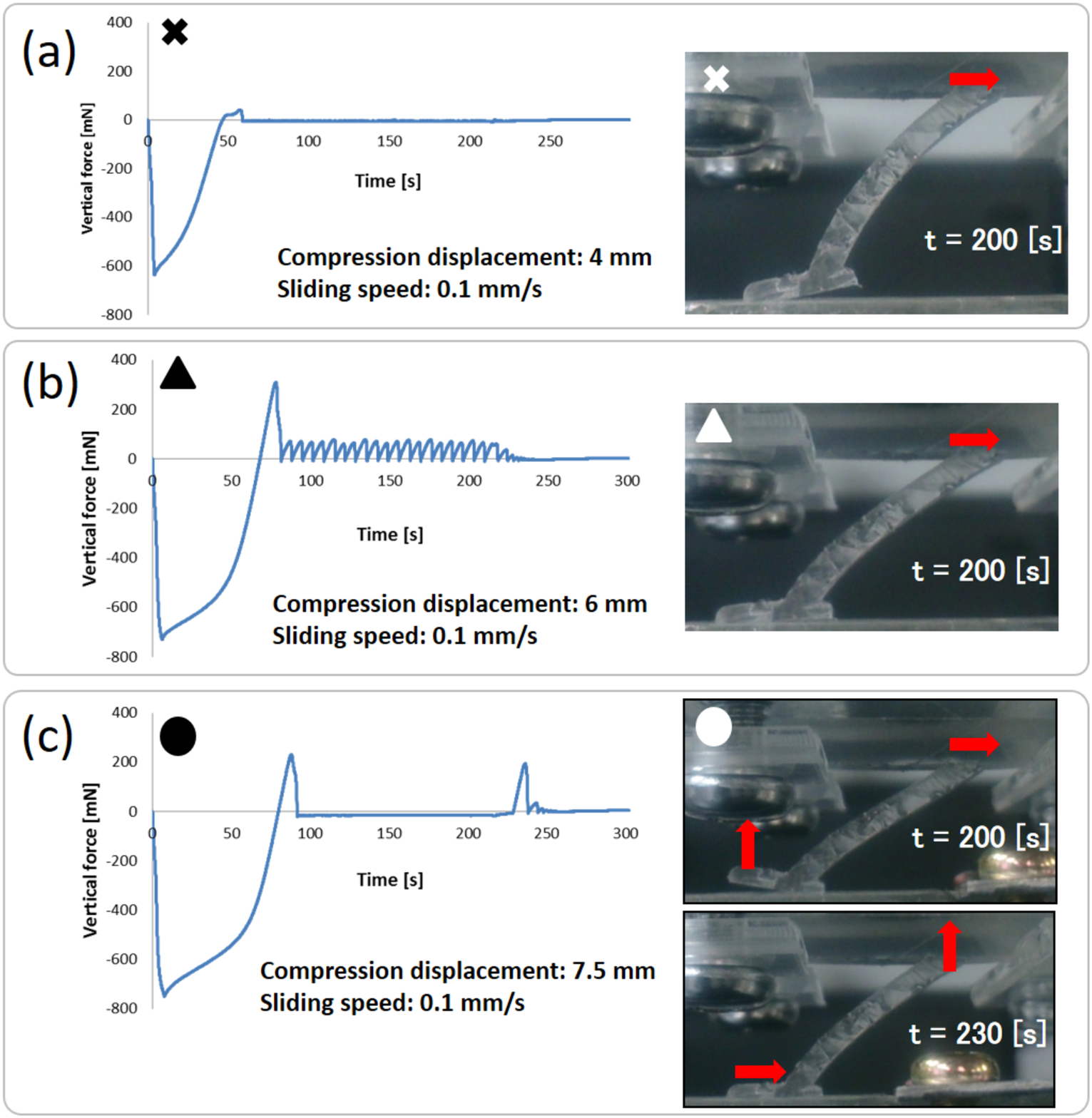}
\caption{Time evolutions of vertical forces for sliding distance of 20 mm 
at sliding speed of 0.1 mm/s. Applied compression displacements are 
(a) 4 mm, (b) 6 mm, and (c) 7.5 mm. The snapshots are the images captured 
at respective times described in the right panel.}
\label{fig4}
\end{figure}

A summary of the results of the second series of experiments 
with excess lateral sliding is depicted in the phase diagram in Fig. 5. 

\begin{figure}
\includegraphics[width = 80mm]{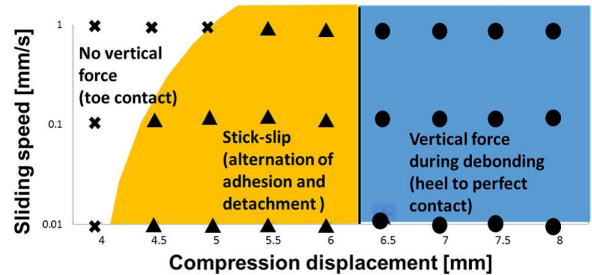}
\caption{Phase diagram of sliding and debonding behaviour 
in second series of experiments. The cross, filled triangle, 
and filled circle symbols correspond to those in Fig. 4.}
\label{fig5}
\end{figure}

Evidently, the compression displacement strongly affects 
the dynamical behaviour of the pad, while the sliding speed affects 
the behaviour only when a small compression displacement is applied. 
Interestingly, the phase boundary between the strong adhesion 
phase (Fig. 4(b)) and the phase with the final contact recovery (Fig. 4(c)) 
is not sensitive to the sliding speed, but depends only on the compression displacement.\\

\section{Conclusion}
In this study, we investigated the effects of the geometry of a gecko-inspired adhesive 
pad on the adhesion/debonding properties. We found that the debonding behaviour 
strongly depends on the direction of lateral sliding, and that such systems are robust 
in terms of excess sliding and vertical compression. 
These results highlight the importance of optimal shape design for individual pillars 
in terms of the adhesion properties for synthetic gecko-like adhesives. 
Although some new findings were obtained through this study, several aspects 
remain to investigated. These include validation of the results of this study 
based on in-situ electron microscope observation of gecko feet, fabrication 
of gecko-inspire adhesives having fine multiple structures 
of adhesive pads attached to a curved beam, 
and optimization of the pad in terms of adhesion performance. Moreover, 
in the adhesion test conducted in this study, only the vertical tensile force 
was measured to evaluate the adhesion performance of the pad. 
It would be worthwhile to measure the tensile and friction forces simultaneously. 
We intend to focus on these topics in future studies.\\

\noindent{\em Acknowledgment}
This work was supported by JSPS KAKENHI grant No.JP25120509 
"Innovative Materials Engineering Based on Biological Diversity". 



\begin{thebibliography}{99}
\bibitem{Maderson}
Maderson P. F. A.: 'Keratinized epidermal derivatives as an aid to climbing in gekkonid lizards', \textit{Nature}, 1964, {\bf 203}, 780-781.
\bibitem{Russel}
Russell A. P.: 'A contribution to the functional morphology of the foot of the tokay, Gekko gecko (Reptilia, Gekkonidae)', \textit{J. Zool. Lond.}, 1975, {\bf 176}, 437-476.
\bibitem{Autumn1}
Autumn K., Liang Y. A., Hsieh S. T., Zesch W., Chan W.-P., Kenny W. T., Fearing R., Full R. J.: 'Adhesive force of a single gecko foot-hair', \textit{Nature}, 2000, {\bf 405}, 681-685. 
\bibitem{Autumn2}
Autumn K., Hsieh S. T., Dudek D. M., Chen J., Chitaphan C., Full R. J.: 'Dynamics of geckos running vertically', \textit{J. Exp. Biol.}, 2006, {\bf 209}, 260-272.
\bibitem{Autumn3}
Autumn K., Dittmore A., Santos D., Spenko M., Cutkosky M.: 'Frictional adhesion: a new angle on gecko attachment', \textit{J. Exp. Biol.}, 2006, {\bf 209}, 3569-3579. 
\bibitem{Hagey}
Hagey T. J., Puthoff J. B., Holbrook M. T., Harmon L. J., Autumn K.: 'Variation in setal micromechanics and performance of two gecko species', \textit{Zoomorphology}, 2014, {\bf 133}, 111-126.
\bibitem{Autumn4}
Autumn K., Sitti M., Peattie A., Hansen W., Sponberg S., Liang Y. A., Kenny T., Fearing R., Israelachvili J., Full R. J.: 
'Evidence for van der Waals adhesion in gecko setae', \textit{Proc. Natl. Acad. Sci.}, 2002, {\bf 99}, 12252-12256.
\bibitem{Pocius}
Pocius A. V.: 'Adhesion and Adhesives Technology' (Carl Hanser Verlag, Munich, 1996).
\bibitem{Ruibal}
Ruibal R., Ernst V.: 'The structure of the digital setae of lizards', \textit{J. Morphol.}, 1965, {\bf 117}, 271-294.
\bibitem{Geim}
Geim A. K., Dubonos S. V., Grigorieva I. V., Novoselov K. S., Zhukov A. A., Shapoval S. Yu.: 'Microfabricated adhesive mimicking gecko foot-hair', {\textit Nature Materials}, 2003, {\bf 2}, 461-463.
\bibitem{Kovalev}
Kovalev A. E., Varenberg M., Gorb S. N.: 'Wet versus dry adhesion of biomimetic mushroom-shaped microstructures', {\textit Soft Matter}, 2012, {\bf 8}, 7560-7566.
\bibitem{Sitti}
Sitti M., Fearing R. S.: 'Synthetic gecko foot-hair micro/nano-structures as dry adhesives', \textit{J. Adhes. Sci. Tech.}, 2003, {\bf 17}, 1055-1073.
\bibitem{Lee}
Lee J., Fearing R. S.: 'Contact Self-Cleaning of Synthetic Gecko Adhesive from Polymer Microfibers', \textit{Langmuir}, 2008, {\bf 24},10587-10591.
\bibitem{Majidi}
Majidi C., Groff R. E., Maeno Y., Schubert B., Baek S., Bush B., Maboudian R., Gravish N., Wilkinson M., Autumn K., Fearing R. S.: 
'High friction from a stiff polymer using microfiber arrays', \textit{Phys. Rev. Lett.}, 2006, {\bf 97}, 076103. 
\bibitem{Ge}
Ge L., Sethi S., Ci L., Ajayan P. M., Dhinojwala A.: 
'Carbon nanotube-based synthetic gecko tapes', \textit{Proc. Natl. Acad. Sci.}, 2007, {\bf 104}, pp.~10792--10795.
\bibitem{Sethi}
Sethi S., Ge L., Ci L., Ajayan P. M., Dhinojwala A.: 'Gecko-inspired carbon nanotube-based self-cleaning adhesives', \textit{Nano letters}, 2008, {\bf 8}, 822-825. 
\bibitem{Maeno}
Maeno Y., Nakayama Y.: 'Geckolike high shear strength by carbon nanotube fiber adhesives', \textit{Appl. Phys. Lett.}, 2009, {\bf 94}, 012103. 
\bibitem{Wang}
Wang L., Hui Y., Fu C., Wang Z., Zhang M., Zhang T.: 'Recent advances in Gecko-inspired adhesive materials and application', \textit{Journal of Adhesion Science and Technology}, 2020, {\bf 0}, 1-17.
\bibitem{Zhao}
Zhao B., Pesika N., Zeng H., Wei Z., Chen Y., Autumn K., Turner K., Israelachvili J.: 
'Role of tilted adhesion fibrils (setae) in the adhesion and locomotion of gecko-like systems', 
\textit{J. Phys. Chem. B}, 2009, {\bf 113}, 3615-3621.
\bibitem{Yamaguchi}
Yamaguchi T., Gravish N., Autumn K., Creton C.: 'Microscopic Modeling of the Dynamics of Frictional Adhesion in the Gecko Attachment System' \textit{J. Phys. Chem. B}, 2009, {\bf 113}, 3622-3628.
\bibitem{Gillies}
Gillies A. G., Fearing R. S.: 'Simulation of synthetic gecko arrays shearing on rough surfaces', \textit{J. R. Soc. Interface}, 2014, {\bf 11}, 20140021. 
\bibitem{Spolenak}
Spolenak R., Gorb S., Arzt E.: 'Adhesion design maps for bio-inspired attachment systems', \textit{Acta Biomater.}, 2005, {\bf 1}, 5-13.
\bibitem{Takata}
Takata M., Yamaguchi T., Gong J. P., Doi M.: 'Electric Field Effect on the Sliding Friction of a Charged Gel', \textit{J. Phys. Soc. Jpn.}, 2009, {\bf 78}, 084602.
\bibitem{Takata2}
Takata M., Yamaguchi T., Doi M.: 'Friction Control of a Gel by Electric Field in Ionic Surfactant Solution', 
\textit{J. Phys. Soc. Jpn.}, 2010, {\bf 79}, 063602. 
\bibitem{Suzuki}
Suzuki R., Yamaguchi T., Doi M.: 'Frictional Property of Hydrogels Prepared under Electric Fields', \textit{J. Phys. Soc. Jpn.}, 2009, {\bf 82}, 124803.
\bibitem{Murakami}
Murakami T., Yarimitsu S., Sakai N., Nakashima K., Yamaguchi T., Sawae Y.: 'Importance of adaptive multimode lubrication mechanism in natural synovial joints', \textit{Tribology International}, 2017, {\bf 113}, 306-315. 
\bibitem{Sano}
T. G. Sano, T. Yamaguchi, H. Wada, \textit{Phys. Rev. Lett.}, 2017, {\bf 118}, 178001. 
\bibitem{Croll}
Croll A. B., Hosseini N., Bartlett M. D.: 'Switchable adhesives for multifunctional interfaces', \textit{Adv. Mater. Technol.}, 2019, {\bf 4}, 1900193.
\bibitem{Murphy}
Murphy M. P., Aksak B., Sitti M.: 'Gecko-inspired directional and controllable adhesion', \textit{Small}, 2009, {\bf 5}, 170-175. 
\bibitem{Gravish}
Gravish N., Wilkinson M., Sponberg S., Parness A., Esparza N., Soto D., Yamaguchi T., Broide M., Cutkosky M., Creton C., Autumn K.: 
'Rate-dependent frictional adhesion in natural and synthetic gecko setae', \textit{J. Roy. Soc. Interface}, 2010, {\bf 7}, 259-269.
\bibitem{Chaudhury}
Chaudhury, M., Kim, K: 'Shear-induced adhesive failure of a rigid slab in contact with a thin confined film', 
\textit{Eur. Phys. J. E}, 2007, {\bf 23}, 175-183.
\bibitem{Yamaguchi2}
Yamaguchi T., Ohmata S., Doi M.: 'Regular to chaotic transition of stick-slip motion in sliding friction of an adhesive gel-sheet', 
\textit{J. Phys.: Condensed Matter}, 2009, {\bf 21}, 205105.
\bibitem{Morishita}
Morishita M, Kobayashi M., Yamaguchi T., Doi M.: 'Observation of spatio-temporal structure in stick{\textendash}slip motion of an adhesive gel sheet', 
\textit{J. Phys.: Condensed Matter}, 2010, {\bf 22}, 365104.
\bibitem{Yamaguchi3}
Yamaguchi T., Sawae Y., Rubinstein S. M.: 'Effects of loading angles on stick-slip dynamics of soft sliders', \textit{Extreme Mechanics Letters}, 2016, {\bf 9}, 331-335.
\bibitem{Tian}
Tian U., Tao D., Pesika N., Wan J., Meng Y., Zhang X.: 'Flexible control and coupling of adhesion and friction of gecko setal array during sliding', \textit{Tribology  Online}, 2015, {\bf 10}, 106-114.
\end{thebibliography}
%


\end{document}